# Growth and characterization of insulating ferromagnetic semiconductor (Al,Fe)Sb

Le Duc Anh[1,a], Daiki Kaneko[1], Pham Nam Hai[2], and Masaaki Tanaka[1,b]

*1. Department of Electrical Engineering and Information Systems,*

*The University of Tokyo, 113-8656 Tokyo, Japan*

*2. Department of Physical Electronics, Tokyo Institute of Technology,*

*2-12-1 Ookayama, Meguro, Tokyo 152-0033, Japan*

We investigate the crystal structure, transport and magnetic properties of Fe-doped ferromagnetic semiconductor $(Al_{1-x},Fe_x)Sb$ thin films up to $x = 14\%$ grown by molecular beam epitaxy. All the samples show p-type conduction at room temperature and insulating behavior at low temperature. The $(Al_{1-x},Fe_x)Sb$ thin films with $x \leq 10\%$ maintain the zinc blende crystal structure of the host material AlSb. The $(Al_{1-x},Fe_x)Sb$ thin film with $x = 10\%$ shows intrinsic ferromagnetism with a Curie temperature ($T_C$) of 40 K. In the $(Al_{1-x},Fe_x)Sb$ thin film with $x = 14\%$, a sudden drop of mobility and $T_C$ was observed, which may be due to microscopic phase separation. The observation of ferromagnetism in (Al,Fe)Sb paves the way to realize a spin-filtering tunnel barrier that is compatible with well-established III-V semiconductor devices.

---

[a] Electronic mail: anh@cryst.t.u-tokyo.ac.jp

[b] Electronic mail: masaaki@ee.t.u-tokyo.ac.jp

Injection of highly-spin-polarized carriers into semiconductors is essential to realize spin-based electronic devices, which are expected to lead to next-generation electronics with low-power consumption. A promising way of spin injection to semiconductors is to use a ferromagnetic insulator as a tunneling barrier, in which the spin-split band structure gives different barrier heights for up-spin and down-spin carriers. Since the tunneling current exponentially decreases with increasing the barrier height, one type of spin is effectively filtered out, leading to a highly-spin-polarized current (the "spin-filtering effect"). The tunneling process is not affected by the conduction mismatch, which is usually a serious problem when directly injecting a spin-polarized current from a ferromagnetic metallic electrode into a semiconductor. So far, however, most of the potential ferromagnetic materials for spin-filtering are Europium chalcogenides (EuO[1], EuS[2], EuSe[3]) or complex oxides (CoFe$_2$O$_4$[4], NiFe$_2$O$_4$[5], NiMn$_2$O$_4$[6], BiMnO$_3$[7], CoCrO$_4$[8] and MnCr$_2$O$_4$[9]), which are not compatible with mainstream semiconductors. Thus, an insulating ferromagnetic material that is compatible with well-established semiconductors is highly needed.

In this paper, we report on the growth and properties of such a material, insulating ferromagnetic semiconductor (FMS) (Al,Fe)Sb. In FMSs, magnetic impurities are doped in conventional semiconductors at a level of several percentages, inducing ferromagnetism while preserving the crystal structure and other important properties of the host semiconductors. Because FMSs are compatible with conventional semiconductors, there were some efforts to realize a FMS-based ferromagnetic tunnel barrier. After the success of Mn-based III-V FMSs such as (Ga,Mn)As and (In,Mn)As, some works have been reported on Mn doped AlAs[10,11] and AlSb[11]. As these materials would work as lattice-matched tunnel barriers in III-V semiconductor heterostructures, spin-filters based on Al-V (V is As or Sb) would significantly broaden the potential of III-V based spin devices such as spin hot-carrier transistors[12], and thus are highly desired. However, Mn-doped AlAs is paramagnetic while there is no experimental report on Mn-doped AlSb. The main difficulty is that a significant amount of carriers are needed to induce ferromagnetism, which cannot be fulfilled in the insulating Mn-doped Al-V.

Recently, Fe-doped III-V FMSs have been successfully grown[13]. Since Fe magnetic impurities are iso-electronic in III-V, one can grow n-type FMS (In,Fe)As[14-18] and p-type FMS (Ga,Fe)Sb[19], both show unexpectedly strong ferromagnetism. Here, by doping Fe in AlSb, we found that a $(Al_{1-x},Fe_x)Sb$ sample with an Fe content $x$ of 10% exhibits intrinsic ferromagnetism while being insulating at low temperature, and thus can be used as a spin filter. As InAs, GaSb, and AlSb are in a lattice-matched group, in which AlSb has the widest bandgap (direct bandgap of 2.3 eV and indirect bandgap of 1.6 eV at room temperature) and acts as a potential barrier, this result provides a complete set of Fe-based FMSs family for spin device applications.

The (Al,Fe)Sb thin films were grown by low-temperature molecular beam epitaxy (LT-MBE) on semi-insulating GaAs (001) substrates. Figure 1(a) illustrates the sample structure. First, we grew a GaAs buffer layer at a substrate temperature of 570°C to smoothen the surface. Next, we grew a 25 nm-thick AlSb (or GaSb) buffer layer at 550°C (or 500°C). The growth of AlSb (GaSb) at a relatively high temperature helps relax quickly the lattice mismatch between AlSb (GaSb) and GaAs. Then we decreased the substrate temperature to 470°C (430°C for GaSb) and continued growing the AlSb (GaSb) buffer up to 100 nm in thickness. After the sample was cooled to $T_S$ (= 236 ~ 280°C), we grew a $(Al_{1-x},Fe_x)Sb$ layer (thickness $t$ = 30, 100 nm) and a 5 nm-thick GaSb cap layer. Most of the samples were grown on an AlSb buffer, except three samples grown on a GaSb buffer for X-ray diffraction (XRD) measurements.

Table I summarizes the parameters $T_S$, $x$, $t$, resistivity $\rho$, hole mobility $\mu$, and hole density $p$ at room temperature of all the (Al,Fe)Sb samples grown on an AlSb buffer studied in this work. The transport properties of these samples were characterized by Hall effect and four-terminal resistivity measurements using 50 (width)×200 (length) μm² Hall bars. All the samples show p-type conduction at room temperature. In Fig. 1(b), we show in the inset the potential energy profile of the valence band edge $E_V$ at the Γ point of these samples, with the hole energy plotted upwards. Here we assume that the band structure of (Al,Fe)Sb is the same as that of AlSb, and the Fermi level $E_F$ is pinned at 0.2 eV above $E_V$ (in the bandgap) at the top GaSb surface[20] and at the mid-gap at the GaAs buffer/GaAs substrate interface. Due to the valence band offsets of 0.2 eV and 0.45 eV at the AlSb/GaAs and GaSb/(Al,Fe)Sb interfaces[21], respectively, holes are confined and transported in the top GaSb (5 nm)/(Al,Fe)Sb (30 nm, red part)/AlSb (100 nm, white part) tri-layers. Thus we used the total thickness of these conducting layers (135 nm) in the estimation of the transport properties. Meanwhile, the magnetic properties of these samples were characterized by magnetic circular dichroism (MCD) and superconducting quantum interference device (SQUID) magnetometry.

First, we studied the three samples with the same $x$ of 8% and different $T_S$ of 236°C, 260°C and 280°C (samples A, B2 and C, respectively). As shown in Table I, sample B2 has the highest mobility $\mu$ and a hole density $p$ of one order of magnitude smaller than that of sample A and C, indicating a good crystal quality with fewer defects. Although all these three samples are paramagnetic, sample B2 shows the largest and a slightly non-linear MCD *vs*. magnetic field characteristics at low temperature. Thus we concluded that $T_S = 260°C$ is the best growth temperature for (Al,Fe)Sb.

Next, we kept $T_S = 260°C$ and investigated the $x$-dependence of the crystal structure, transport and magnetic properties of $(Al_{1-x},Fe_x)Sb$ thin films. We prepared samples B0 – B4 with $x$ of 0, 2, 8, 10, 14%, respectively, in which sample B0 is a reference sample. Figure 1(c) shows *in situ* reflection high energy electron diffraction (RHEED) patterns during the MBE growth of the 30 nm-thick (Al,Fe)Sb layer of samples B1 - B4. The RHEED patterns of samples B1 - B3 ($x$ = 2 - 10%) were bright and relatively streaky, but turned spotty in sample B4 ($x$ = 14%). All the RHEED patterns reflect the zinc blende (ZB) crystal structure. Figure 1(d) shows transmission electron microscopy (TEM) images of sample B3 ($x$ = 10%). The left panel of Fig. 1(d) shows the cross sectional TEM image of the GaSb/(Al,Fe)Sb/AlSb layers. The upper and lower panels on the right are high-resolution lattice images of the (Al,Fe)Sb layer near the surface (upper panel) and AlSb buffer (lower panel) measured by scanning TEM. These TEM images indicate the ZB crystal structure of the (Al,Fe)Sb layer, which is also confirmed by the transmission electron diffraction (TED) pattern as shown in the inset of the left panel. These results show that (Al,Fe)Sb with the Fe concentration up to 10% maintains the ZB crystal structure without any second phase.

Figure 2(a) shows XRD results of the (Al$_{1-x}$,Fe$_x$)Sb samples with $x$ = 5, 10 and 14%. The (Al,Fe)Sb layers in these samples were grown at $T_S$ = 260°C on a GaSb buffer to avoid diffractions from the AlSb buffer layers, whose lattice constant is very close to that of (Al,Fe)Sb. All the three samples show a sharp GaAs (004) peak from the GaAs substrate, and a broader peak that can be deconvoluted to two Gaussian curves corresponding to the GaSb (004) peak from the GaSb buffer layer and the (Al,Fe)Sb (004) peak from the (Al,Fe)Sb layer. From the peak positions, we estimated the intrinsic (without strain) lattice constants of (Al,Fe)Sb as shown in Fig. 2(b) (see Supplementary Material (S.M.)). The lattice constant of the (Al$_{1-x}$,Fe$_x$)Sb layers linearly decreases with increasing the Fe content $x$ following the Vegard's law, which indicates that Fe atoms reside in the cation sites of the AlSb lattice.

The transport properties of samples B0 – B4 are summarized in Table I. Samples B0 - B3 ($x$ = 0 - 10%) show p-type conduction with the hole density $p$ around $2 \sim 3 \times 10^{17}$ cm$^{-3}$, which is not correlated with $x$. This indicates that the Fe atoms are in the neutral Fe$^{3+}$ state and do not supply carriers, similar to the behavior observed in (In,Fe)As[14] and (Ga,Fe)Sb[19]. However, in sample B4 ($x$ = 14%), $p$ suddenly increases by two orders of magnitude, to 3.1 $\times 10^{19}$ cm$^{-3}$. The sudden increase of $p$ for $x$ = 14% was reproduced in two samples grown under the same conditions. On the other hand, the mobility $\mu$ increases quickly with $x$ (= 2, 8, 10%), and reaches the highest value of 49.4 cm$^2$/Vs in sample B3 ($x$ = 10%), then suddenly drops to 0.5 cm$^2$/Vs in sample B4 ($x$ = 14%). This result indicates that there is a significant change in the crystal quality when increasing $x$ from 10 to 14%. Nevertheless, all the samples show insulating behavior at low temperature. The temperature dependence of $\rho$ and $p$ of sample B3 ($x$ = 10%) are representatively shown in Fig. 1(b). The resistivity $\rho$ increased by three orders of magnitude when the sample was cooled from 300 K to 3.5 K. Because of the high resistivity at low temperatures, the Hall effect can only be measured at $T$ > 100 K. The hole density $p$ decreases from $10^{17}$ to $10^{16}$ cm$^{-3}$ with decreasing temperature from 300 K to 100 K, that is a typical behavior of a non-degenerated semiconductor.

Figure 3(a) shows the MCD spectra of samples B0 - B4, measured at 5 K under a magnetic field of 1 T applied perpendicular to the film plane. For sample B0 ($x = 0\%$) we observed two peaks at 2.39 eV and 2.89 eV, corresponding to the critical point energies $E_0$ and $E_1$ of AlSb at low temperature (the $E_0$ and $E_1$ energies of AlSb at room temperature are 2.27 eV and 2.83 eV, respectively[22]). From sample B0 to B3 ($x = 0 - 10\%$), the MCD intensity of the $E_0$ and $E_1$ peaks increases quickly, indicating that the Zeeman splitting of the (Al,Fe)Sb band structure is largely enhanced with increasing $x$ up to 10%. The MCD spectra of samples B0 - B3 indicate that the $(Al_{1-x},Fe_x)Sb$ layers with $x \leq 10\%$ maintain the ZB crystal and band structure of the AlSb host material, consistent with the observation by TEM. On the other hand, sample B4 ($x = 14\%$) shows a different MCD spectrum with smaller but much broader peaks from 1.5 eV to 3.3 eV. This indicates the possibility of microscopic phase separation which cannot be seen from the RHEED and XRD results. Figure 3(b) shows the magnetic field dependence of the MCD intensity (MCD-$H$ curves) at the $E_1$ peak of all the samples B1 - B4. Non-linear hysteresis MCD-$H$ curves were observed for the samples with $x \geq 8\%$. From the Arrott plots of the MCD-$H$ curves, we estimated the Curie temperature ($T_C$) of all the samples as shown in Fig. 3(c), together with their hole density $p$. The (Al,Fe)Sb samples with $x \geq 10\%$ exhibit ferromagnetism, with the highest $T_C = 40$ K at $x = 10\%$. A sudden drop of $T_C$ to 10 K was observed in sample B4 ($x = 14\%$).

We further investigated the magnetic properties of the ferromagnetic sample B3 ($x = 10\%$) with $T_C = 40$K. Figures 4(a) and (b) show the MCD-$H$ curves of sample B3 measured at the $E_1$ peak at various temperatures and their Arrott plot, respectively, which indicates $T_C$ = 40 K. In Fig. 4(c) we show the field cooling curve (FC) and the zero field cooling curve (ZFC) of sample B3, which are the magnetization measured by SQUID when increasing temperature from 10 K to 300 K (*M-T* curves). Before the *M-T* measurements, the sample was cooled to 10 K with a magnetic field of 1 T applied perpendicular to the film plane in the FC measurement, while the magnetic field was absent in the cooling process of the ZFC measurement. The monotonic behavior of both FC and ZFC curves confirms the absence of ferromagnetic nanoclusters, which, if existed, would show a blocking temperature due to superparamagnetism. Note that because of the small remanent magnetization of sample B3, we had to apply a weak perpendicular magnetic field of 50 Oe during these *M-T* measurements which caused a tail up to temperatures above $T_C$. However, the inverse of the magnetic susceptibility $\chi^{-1} = H/M$ (at $H$ = 50 Oe, $T$ > 100 K, green open circles in Fig. 4(c)), which follows the Curie-Weiss law $\chi = C/(T-T_C)$ (green dotted line, $C$ is the Curie constant), indicates $T_C$ = 40 K. This $T_C$ agrees with the value estimated by the Arrott plot of the MCD-*H* characteristics. Finally, as shown in Fig. 4(d), the magnetization curves of sample B3 measured at 10 K with an perpendicular magnetic field measured by SQUID and MCD (at the $E_1$ peak) show good agreement with each other, indicating a single source of ferromagnetism in the (Al,Fe)Sb layer. The saturated magnetic moment per Fe atom was estimated to be 1.83 $\mu_B$.

One can see from Fig. 3(c) that $T_C$ of the (Al,Fe)Sb samples is not proportional to $p$ unlike typical FMSs. This indicates that the observed ferromagnetism in (Al,Fe)Sb is not carrier-induced. With such a low hole density ($< 10^{16}$ cm$^{-3}$) at low temperatures, long-range exchange interactions between the Fe magnetic moments cannot be mediated effectively by holes. In this case, short-range exchange interactions between Fe atoms may play an essential role. Here we suggest a possible scenario of short-range exchange interactions. Considering the low solubility of Fe in III-V semiconductors, spinodal decomposition may occur and cause local fluctuation in the Fe concentration while preserving the ZB crystal structure of the host material, as observed in many FMSs[14,23,24]. In local Fe-rich domains, Fe atoms can sit in the second-nearest-neighbor sites and interact with each other through mechanisms such as superexchange interaction which is effective even at low carrier density. As illustrated in Fig. 4(e), these Fe-rich domains coexist with paramagnetic Fe-poor areas, and would behave as discrete ferromagnetic domains. At low magnetic field, their magnetizations point to different directions, which explains the small remanent magnetization and the small effective magnetic moment per Fe atom of the (Al,Fe)Sb samples. Such a discrete ferromagnetic domain structure has been observed in (In,Fe)As[14]. Considering the ratio of the effective magnetic moment of 1.83 $\mu_B$ per Fe atom in sample B3 to the expected value of the Fe$^{3+}$ ion (5$\mu_B$), we estimate that only 36.6% of the Fe atoms are contributing to the ferromagnetism.

In summary, we have grown an insulating ferromagnetic semiconductor (Al,Fe)Sb by LT-MBE. The (Al$_{1-x}$,Fe$_x$)Sb layers maintain the ZB crystal structure up to the Fe content $x$ = 10%, and show insulating behavior at low temperature. The observation of intrinsic ferromagnetism in insulating (Al,Fe)Sb paves the way to realize spin-filtering tunnel barrier in III-V semiconductor spin devices.

**Acknowledgments**

This work is supported by Grant-in-Aids for Scientific Research including the Specially Promoted Research, the Project for Developing Innovation Systems of MEXT and the Cooperative Research Project Program of RIEC, Tohoku University. L. D. A. acknowledges the JSPS Fellowship for Young Scientists (KAKENHI number 257388) and the Program for Leading Graduate Schools (MERIT). P. N. H. acknowledges the Murata Science Foundation, the Yazaki Foundation for Science and Technology, and the Toray Science Foundation.

**Figure captions**

FIG 1. (a) Schematic sample structure. (b) Temperature dependence of the resistivity $\rho$ (black curve, left axis) and the hole density $p$ (red circles, right axis) of sample B3. The inset illustrates the valence band profile of samples B0 - B4, where the numbers are the band offset values in eV and hole energy plotted upwards. (c) *In situ* RHEED patterns during the MBE growth of the (Al,Fe)Sb layers in samples B1 - B4. (d) Left panel: Cross-sectional TEM image of sample B3 ($x$ = 10%); inset is the TED pattern of the (Al,Fe)Sb layer. Right

panel: High-resolution scanning TEM image of the area close to the surface (top panel) and to the AlSb buffer (bottom panel).

FIG 2. (a) XRD rocking curves of the $(Al_{1-x},Fe_x)Sb$ samples with $x$ of 5, 10 and 14%, respectively. The broad peak was fitted by two Gaussian curves corresponding to GaSb (004) from the GaSb layer (black dotted lines) and (Al,Fe)Sb (004) from the (Al,Fe)Sb layer (green dotted lines). For each sample, the sum of the two Gaussian curves are plotted by the red dotted curve. (b) Estimated intrinsic lattice constant of $(Al_{1-x},Fe_x)Sb$ as a function of $x$.

FIG 3. (a) MCD spectra of samples B0 - B4, measured at 5 K under a magnetic field of 1T perpendicular to the film plane. (b) Magnetic field dependence of the MCD intensity measured at the photon energy of the $E_1$ peak of samples B0 - B4. (c) $T_C$ (red circles, left axis) and hole density $p$ (blue diamonds, right axis) of samples B0 - B4, plotted as a function of the Fe concentration.

FIG 4. (a) MCD-$H$ curves of sample B3 at various temperatures, measured at the photon energy of the $E_1$ peak . (b) Arrott plot of the MCD - $H$ curves from which $T_C$ is estimated to be 40K. (c) FC (red circles) and ZFC (blue diamonds) $M$-$T$ curves of sample B3. Open green circles are the inverse of the magnetic susceptibility $\chi^{-1}$ = $H/M$, which follows the Curie-Weiss law $\chi = C/(T-T_C)$ (green dotted line) indicating $T_C$ = 40 K. (d) $M$-$H$ curve of sample B3 measured by SQUID (blue diamonds) and MCD (red circles, normalized) at 10K with a perpendicular magnetic field. (e) Illustration of the possible coexistence of discrete ferromagnetic domains formed by Fe-rich domains (pink areas) and paramagnetic domains (white areas) in (Al,Fe)Sb thin films, without (left panel) and with (right panel) external

magnetic field. Small arrows (black and red) represent the Fe local spins, and the large blue arrows represent the magnetization of the domains.

Table I. Growth parameters and transport properties of all the (Al,Fe)Sb samples grown on an AlSb buffer. These transport properties were measured by the Hall effect measurements and the four-terminal resistivity measurements at room temperature using $50 \times 200$ μm$^2$ Hall bars.

| Sample | Ts (°C) | Fe content $x$ (%) | (Al,Fe)Sb thickness $t$ (nm) | $\rho$ ($\Omega$ cm) | $\mu$ (cm$^2$/Vs) | Hole density $p$ (cm$^{-3}$) |
|---|---|---|---|---|---|---|
| A | 236 | 8 | 100 | 0.33 | 8.6 | $2.2 \times 10^{18}$ |
| C | 280 | 8 | 30 | 0.34 | 16.4 | $1.1 \times 10^{18}$ |
| B0 | 260 | 0 | 30 | 0.37 | 46.9 | $3.5 \times 10^{17}$ |
| B1 | 260 | 2 | 30 | 4.9 | 4.73 | $2.6 \times 10^{17}$ |
| B2 | 260 | 8 | 30 | 2.0 | 16.9 | $1.9 \times 10^{17}$ |
| B3 | 260 | 10 | 30 | 0.38 | 49.4 | $3.2 \times 10^{17}$ |
| B4 | 260 | 14 | 30 | 0.4 | 0.51 | $3.1 \times 10^{19}$ |

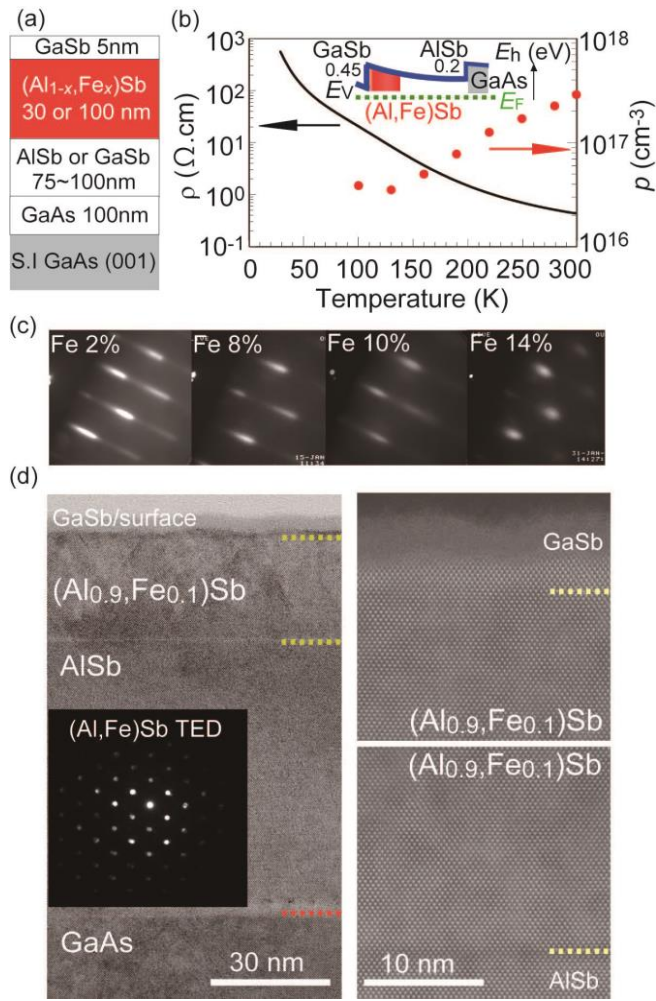

**FIG 1.** Anh *et al.*

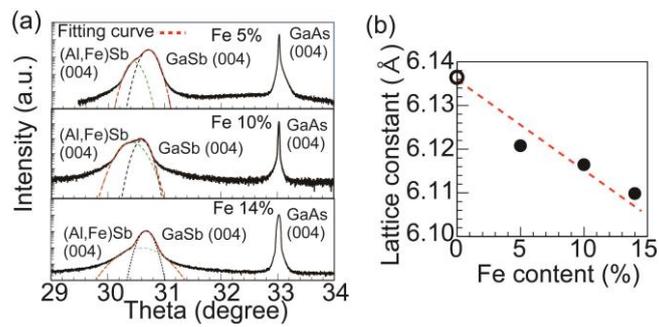

**FIG 2.** Anh *et al.*

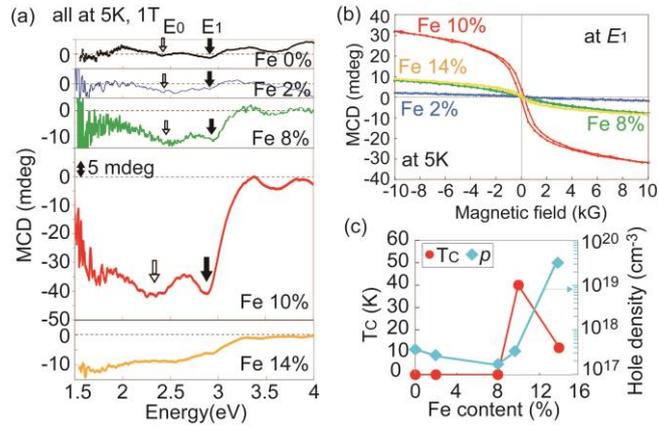

**FIG 3.** Anh *et al.*

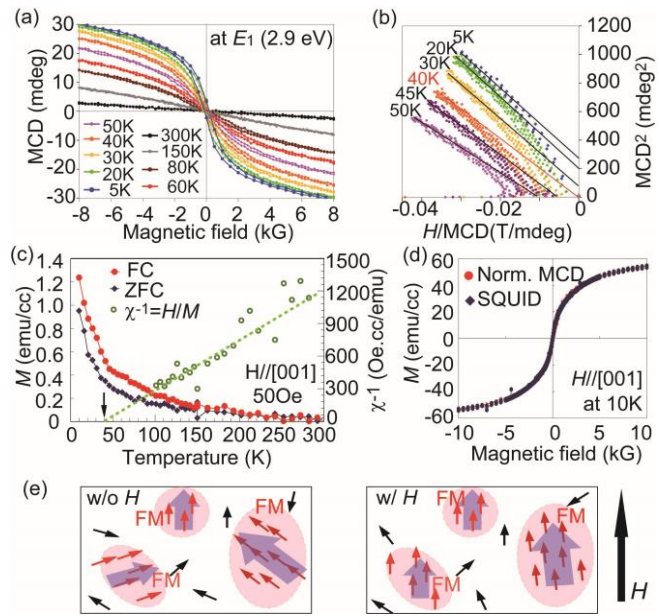

**FIG 4.** Anh *et al.*

# Supplemental Material

# Growth and characterization of ferromagnetic insulator (Al,Fe)Sb

Le Duc Anh[1], Daiki Kaneko[1], Pham Nam Hai[2], and Masaaki Tanaka[1]

*1. Department of Electrical Engineering and Information Systems,
The University of Tokyo, 113-8656 Tokyo, Japan
2. Department of Physical Electronics, Tokyo Institute of Technology,
2-12-1 Ookayama, Meguro, Tokyo 152-0033, Japan*

1. **Reflection magnetic circular dichroism (MCD) and its Arrott plot**

    In reflection MCD, we measure the difference in optical reflectivity between right ($R_{\sigma+}$) and left ($R_{\sigma-}$) circular polarizations, that is induced by the spin splitting of the band structure, at a magnetic field $H$ of 1 T applied perpendicular to the film plane. The MCD intensity is expressed by $\mathrm{MCD} = \frac{90}{\pi}\frac{(R_{\sigma+} - R_{\sigma-})}{2R} \propto \frac{90}{\pi}\frac{1}{2R}\frac{dR}{dE}\Delta E$, where $R$ is the optical reflectivity, $E$ is the photon energy, and $\Delta E$ is the spin-splitting energy (Zeeman energy) of the material. Because of the difference of the $dR/dE$ term, a MCD spectrum shows peaks corresponding to the optical transitions at critical point energies of the FMS band structure. At the same time, the MCD intensity is proportional to the magnetization $M$ of a material, because $M \propto \Delta E$ in FMSs. Therefore, MCD measurements give information of both the magnetization and the electronic structure of the material. Magnetization characteristics of all the samples were measured by magnetic field dependence of the MCD intensity at the $E_1$ peak of (Al,Fe)Sb. $T_C$ of all the samples were estimated using the Arrott plot[1], which is based on the $\mathrm{MCD}^2 - H/\mathrm{MCD}$ plots at different temperatures.

2. **Lattice constant of (Al,Fe)Sb estimated by XRD measurements.**

    We measured XRD of (Al,Fe)Sb samples by a Rigaku's Smart-Lab® system with a copper source at a X-ray wavelength of 1.54 Å. By fitting the XRD peaks of GaSb and (Al,Fe)Sb using Gaussian curves, we determined the peak position (the $\theta$ value), and estimated the strained lattice constant $a_{z,\mathrm{GaSb}}$ of GaSb and $a_{z,\mathrm{(Al,Fe)Sb}}$ of (Al,Fe)Sb along the growth direction ($z$ axis = [001]). The strain $e_z$ along the $z$-axis of the GaSb buffers was calculated by $e_z = (a_{z,\mathrm{GaSb}} - a_{i,\mathrm{GaSb}})/a_{i,\mathrm{GaSb}}$, where $a_{i,\mathrm{GaSb}}$ is the intrinsic lattice constant of cubic GaSb (0.609593 nm)[2]. Then, using the second order elastic moduli $C_{11,\mathrm{GaSb}}$, $C_{12,\mathrm{GaSb}}$ of

GaSb[3], we calculated the in-plane strain $e_x$ of the GaSb layer as $e_x = -2(C_{11,\text{GaSb}}/C_{12,\text{GaSb}})e_z$. From $e_x$ we estimated the in-plane lattice constant $a_{x,\text{GaSb}}$ of the strained GaSb buffer layers using the relation $e_x = (a_{x,\text{GaSb}} - a_{i,\text{GaSb}})/a_{i,\text{GaSb}}$.

For the (Al,Fe)Sb layers, we assumed that the in-plane lattice constant $a_{x,(\text{Al,Fe})\text{Sb}}$ is the same as that of the GaSb buffer ($a_{x,\text{GaSb}}$). Then, we obtained the intrinsic lattice constant $a_{i,(\text{Al,Fe})\text{Sb}}$ of (Al,Fe)Sb, with all the strain effects excluded, as follows:

$$a_{i,(\text{Al,Fe})\text{Sb}} = \frac{a_{z,(\text{Al,Fe})\text{Sb}} + 2\dfrac{C_{12,\text{AlSb}}}{C_{11,\text{AlSb}}}a_{x,(\text{Al,Fe})\text{Sb}}}{1 + 2\dfrac{C_{12,\text{AlSb}}}{C_{11,\text{AlSb}}}} \qquad (1)$$

Here, we assumed that (Al,Fe)Sb has the same elastic moduli as AlSb, and used the parameters $C_{11,\text{AlSb}}$, $C_{12,\text{AlSb}}$ of AlSb[4,5] in eq. (1). The obtained values of $a_{i,(\text{Al,Fe})\text{Sb}}$ are shown in Fig. 2(b) in the main manuscript.